\documentclass[10pt]{article}
\usepackage{opex2}
\usepackage{epsfig}
\usepackage{amsmath}
\usepackage{amssymb}

\begin{document}

\title{Near-field to far-field transition of photonic crystal fibers: symmetries and interference phenomena}

\author{Niels Asger Mortensen \& Jacob Riis Folkenberg}
\address{Crystal Fibre A/S, Blokken 84, DK-3460 Birker\o d, Denmark}
\email{nam@crystal-fibre.com\\http://www.crystal-fibre.com}

\begin{abstract}

The transition from the near to the far field of the fundamental mode radiating out of a photonic crystal fiber is
investigated experimentally and theoretically. It is observed that the hexagonal shape of the near field rotates two times
by $\pi/6$ when moving into the far field, and eventually six satellites form around a nearly gaussian far-field pattern.
A semi-empirical model is proposed, based on describing the near field as a sum of seven gaussian distributions, which
qualitatively explains all the observed phenomena and quantitatively predicts the relative intensity of the six satellites
in the far field.

\end{abstract}

\ocis{(060.2430) Fibers, single-mode; (230.3990) Microstructure devices; (000.4430) Numerical approximation and analysis}

\section{Introduction}

Photonic crystal fibers (PCF) are a new class of optical fibers which has revealed many surprising phenomena and also holds
a big promise for future applications (see {\it e.g.}~\cite{opticsexpress,joa,knight2002}). These PCFs are made from pure
silica with a cladding consisting of a regular lattice of air-holes running along the fiber axis. Depending on the
arrangement of the air-holes the guiding of light can be provided by either modified total internal
reflection~\cite{knight1996,knight1997errata} or by the photonic band-gap effect~\cite{knight1998,cregan1999} and PCFs
can even be endlessly single-mode~\cite{birks1997} because of the wavelength dependence of the cladding index.
For the basic operation we refer to the review of Broeng {\it et al.}~\cite{broeng1999}.

\begin{figure}[t!]
\begin{center}
\epsfig{file=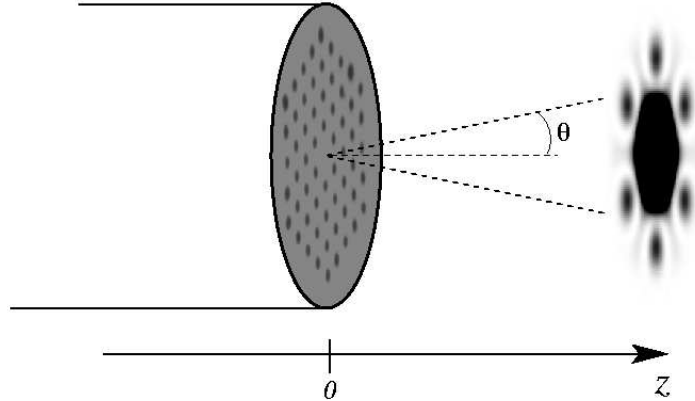, width=0.7\textwidth,clip}
\end{center}
\caption{Schematic of a single-mode PCF ($z<0$) with an end-facet from where light is radiated into free space ($z>0$). }
\label{fig1}
\end{figure}

Understanding the shape and radiation pattern, as illustrated in Fig.~\ref{fig1}, of the mode in the endlessly single-mode
PCF is very important. E.g. in
tests and applications this is essential for estimations of coupling efficiencies and for determining the mode field diameter
from the far-field distribution. Furthermore, it is fundamentally the simplest structure with a hexagonal cladding, and hence
the understanding of this structure will be a natural basis for understanding the modes of more sophisticated PCF structures.
In this paper we present a semi-empirical model which is capable of explaining both the near and far-field
distribution of the mode, but most importantly also accounts for the fine structure in the transition from the near to the
far field. The simplicity of the model allows for a phenomenological interpretation of the shapes of the near and far-field
patterns.

\begin{figure*}[t!]
\begin{center}
\epsfig{file=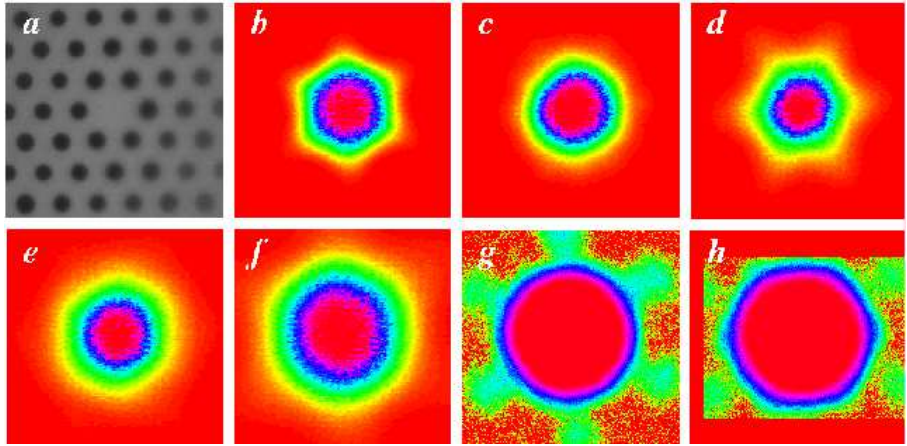, width=0.9\textwidth,clip}
\end{center}
\caption{Experimentally observed near-field intensity distributions for a PCF with $\Lambda\simeq 3.5 \,{\rm \mu m}$ and $d/\Lambda\simeq 0.5$ (micro-graph in panel a) at a free-space wavelength $\lambda=635\,{\rm nm}$. The distance from the end-facet varies from $z=0$ to $z\sim 10\,{\rm \mu m}$ (panels b to f). At a further distance the six low-intensity satellite spots develop (panels g and h, logarithmic scale).  }
\label{fig2}
\end{figure*}

\begin{figure*}[b!]
\begin{center}
\epsfig{file=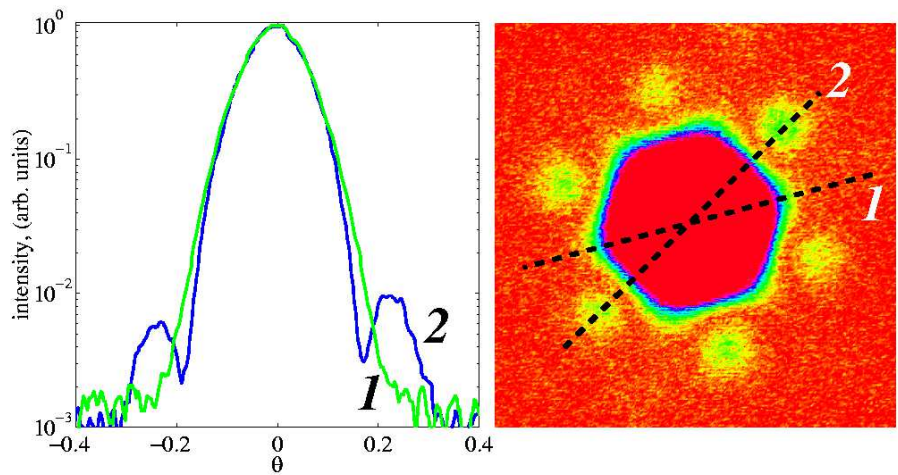, width=0.9\textwidth,clip}
\end{center}
\caption{Experimentally observed far-field intensity distribution showing an overall gaussian profile with
six additional low-intensity satellite spots along one of the two principal directions (line 2). Angles are given in radians.}
\label{fig3}
\end{figure*}

\section{Experiments} 

The measurements reported are for a PCF with a triangular air-hole lattice with pitch of $\Lambda\simeq 3.5 \,{\rm \mu m}$ and
air holes of diameter $d/\Lambda\simeq 0.5$. The measurements reported here were performed at a free-space 
wavelength of $\lambda=635\,{\rm nm}$, where the light is guided in a single mode in the silica core of the fiber
formed by a ``missing'' air hole. In panel a of Fig.~\ref{fig2} a micro-graph of the fiber structure can be seen.

The near-field distribution was measured using a microscope objective to magnify the mode onto a Si-based CCD camera. In
Fig.~\ref{fig2}b the intensity distribution is shown at focus. By translating the fiber away from
the focal plane, the intensity distribution may be imaged at different distances between the near and the far field. This is
shown in panels b to h in Fig.~\ref{fig2}. As expected the mode at focus has a hexagonal shape, that extends in the six regions between
the inner holes and is sharply confined at the six silica-hole interfaces. However, when the image is defocused, the shape
at first transforms into a nearly circular shape (panel c) followed by a hexagonal shape rotated by an angle of $\pi/6$ with respect to the
focus (panel d). After this the shape again becomes close to circular (panel e) , and finally transforms into the original hexagonal orientation (panel f) with six
satellites emerging from the distribution (panels g and h). It is noted that the orientation of the satellites is rotated by
$\pi/6$ with respect to the six inner holes surrounding the core. In Fig.~\ref{fig3} (right) the intensity distribution
in the far-field limit is shown (several centimeters from the fiber end-facet), obtained using a commercial far-field 
profiler. Here, the satellites have fully developed and as shown in the cross sectional plot in Fig.~\ref{fig3}
(left) the peak intensities of the satellites are more than two orders of magnitude lower than the main peak. Hence, a
reasonably accurate analysis of the far field may be performed considering only the main peak.

Apart from being a fascinating and intriguing evolution of the mode shape from the near to the far field, it is important
to be aware of these transitions
in any application that involves imaging of the modes. {\it E.g.} for estimations of the mode field diameter and effective area
based on near-field analysis, it is important to focus the mode correctly, and the positions corresponding to panel b and panel d in Fig.~\ref{fig2} may easily be confused. They both show the hexagonal shape and
have only slightly different mode sizes. Hence, as a measurement procedure for determining the mode field diameter, a direct
measurement of the near field may be even more tricky than it is for ``standard technology fibers'' with circular symmetry.

In panel a of Fig.~\ref{fig4} two cross-sections of the measured near-field distribution are shown, one taken along a line passing
through opposite hole centers (1) and the second taken along a line passing between the holes (2) (rotated by an angle $\pi/6$ with
respect to the first). It is noted that a gaussian distribution is a significantly better fit to the intensity along line
(2), motivating a simple interpretation of the mode shape : the mode is a circular gaussian distribution from which a
narrow distribution localized at each of the centers of the six inner holes is subtracted. This simple interpretation is
theoretically modeled in the following.

\section{Theory} 

In order to simulate the radiated field we start from the fully-vectorial fields in the single-mode all-dielectric PCF
\begin{equation}\label{H_PCF}
{\boldsymbol H}(x,y,z)=\boldsymbol{h}(x,y)\,  e^{\pm i \beta(\omega) z},
\end{equation}
where ${\boldsymbol h}(x,y)$ and $\beta(\omega)$ are the transverse fields and the propagation constant, respectively.
These we calculate numerically by a fully-vectorial plane-wave method~\cite{johnson2000}.

\begin{figure*}[t!]
\begin{center}
\epsfig{file=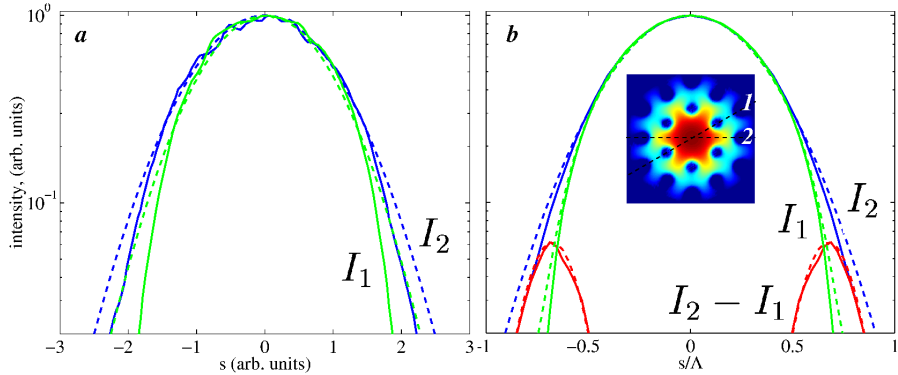, width=0.9\textwidth,clip}
\end{center}
\caption{Panel a shows the experimentally observed near-field intensity along the two principal directions 1 and 2 (see insert of panel b). Panel b shows the numerically calculated intensity distribution in a corresponding ideal PCF with the solid lines showing the intensity along the principal directions and the difference. The blue and red dashed lines show gaussian fits to $I_2$ and $I_2-I_1$ and the dashed green line shows their difference.}
\label{fig4}
\end{figure*}

Substantial insight in the physics of the radiation problem can be gained by expanding ${\boldsymbol h}(x,y)$ in gaussians.
Introducing the notation ${\boldsymbol s}= (x,y)$ and using that the components of ${\boldsymbol h}(x,y)$ can be chosen
either real or imaginary we consider
\begin{equation}
I({\boldsymbol s})=|{\boldsymbol h}({\boldsymbol s})|^2= \Big|\sum_j A_j u({\boldsymbol s}-{\boldsymbol s}_j,w_j)\Big|^2,\; u({\boldsymbol s},w)=\exp(-s^2/w^2).
\end{equation}
For the radiation into free space this gives a linear combination of expanding gaussian beams and this is a well-studied
problem, see {\it e.g.} \cite{ghatak1998,ghatak1989}. Neglecting the small back-scattering from the end-facet, the gaussian
$u({\boldsymbol s},w)$ at finite $z>0$ transforms as

\begin{equation}
u({\boldsymbol s},w)\longrightarrow u({\boldsymbol s},z,w)=\Big(1-i\tfrac{2z}{kw^2}\Big)^{-1}\exp\Big[-ik \Big(z+ \tfrac{s^2}{2R(z)}\Big)-\tfrac{s^2}{W^2(z)}\Big],
\end{equation}
where $R(z)=z(1+k^2w^4/4z^2)$ and $W(z)=w(1+4z^2/k^2w^4)^{1/2}$. In the following we consider a particular simple linear
combination in the PCF;

\begin{figure*}[t!]
\begin{center}
\epsfig{file=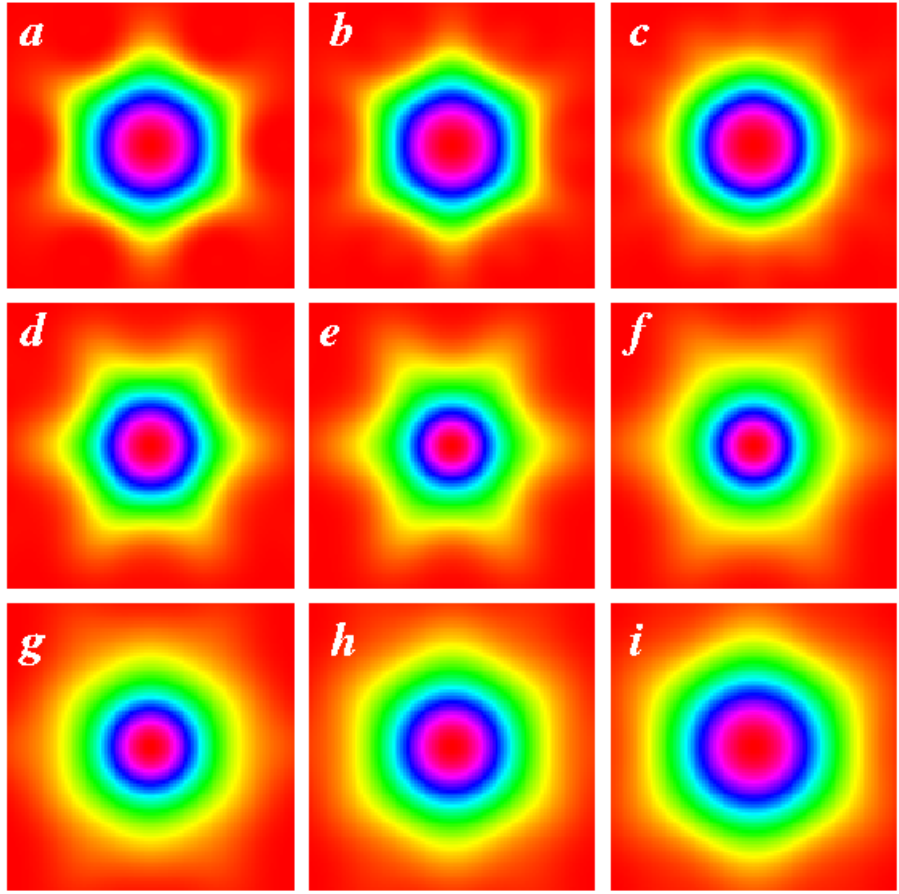, width=0.9\textwidth,clip}
\end{center}
\caption{Near-field intensity distribution calculated from Eq.~(\ref{H_gaussian}) with values of $w_h$, $w_c$, and $\gamma$ determined from the intensity in the PCF obtained by a fully-vectorial calculation, see Fig.~\ref{fig4}. The distance varies from $z=0$ to $z=8\Lambda$ (panels a to i) in steps of $\Delta z= \Lambda$ (see also animation with $\Delta z= \Lambda/4$, {\it http://www.crystal-fibre.com/technology/movie.gif}, 3 Mbyte). }
\label{fig5}
\end{figure*}

\begin{equation}
I({\boldsymbol s})= A^2
\Big| u({\boldsymbol s},w_c)- \gamma \sum_{j=1}^6 u({\boldsymbol s}-{\boldsymbol s}_j,w_h)\Big|^2,
\end{equation}
where ${\boldsymbol s}_j=R_c\times(\cos \tfrac{j 2\pi}{6},\sin\tfrac{j 2\pi}{6})$ with $(\Lambda/R_c){\boldsymbol s}_j $
being the center position of the six air holes nearest to the core. Here, $R_c \sim \Lambda$ (the radius of the silica
core), $w_c\sim \Lambda$ (the mode-field radius), and $w_h \sim d/2$ (the radius of the air holes). The first term gives
the over-all gaussian intensity profile of the mode and with $\gamma \sim u(R_c,w_c)$ the additional six terms of opposite
sign suppress the intensity at the six air-holes nearest to the core. For finite $z>0$ the intensity transforms
as

\begin{equation}\label{H_gaussian}
I({\boldsymbol s})\longrightarrow I({\boldsymbol s},z)= A^2
\Big| u({\boldsymbol s},z,w_c)- \gamma \sum_{j=1}^6 u({\boldsymbol s}-{\boldsymbol s}_j,z,w_h)\Big|^2.
\end{equation}
In panel b of Fig.~\ref{fig4} we show an example of the intensity distribution in an ideal PCF with $\Lambda = 3.5\,{\rm \mu m}$ and
$d/\Lambda=0.5$ at $\lambda=635\,{\rm nm}$ corresponding to experimental situation. For the dielectric function we have used $\varepsilon=1$ for the air holes
and for the silica we have used $\varepsilon = (1.4572)^2=2.123$ based on the Sellmeier formula. While
Eq.~(\ref{H_gaussian}) may seem too simplistic the good fits to gaussians strongly justify it and as we shall see it
reproduces the physics observed experimentally.

\begin{figure*}[bt!]
\begin{center}
\epsfig{file=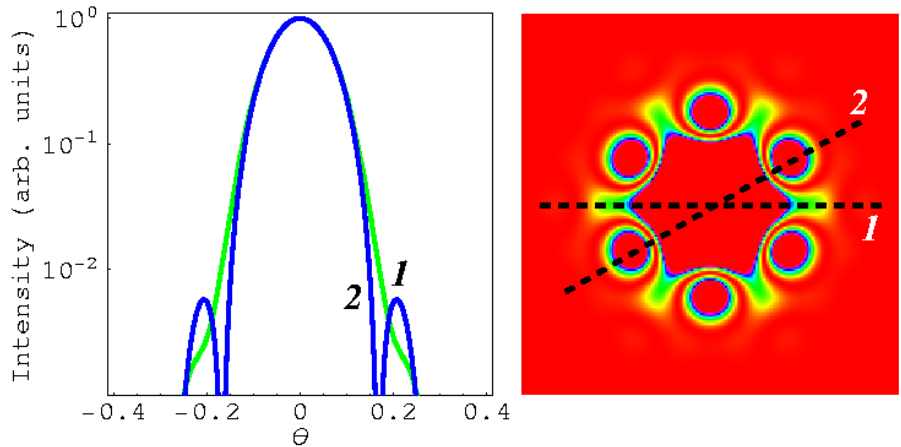, width=0.9\textwidth,clip}
\end{center}
\caption{Far-field intensity distribution ($z=1000\Lambda\gg \lambda$) corresponding to the near field in Fig.~\ref{fig5}. The intensity distribution has an overall gaussian profile with six additional low-intensity satellite spots along one of the two principal directions (line 2).}
\label{fig6}
\end{figure*}

In Fig.~\ref{fig5} we show the corresponding near field based on Eq.~(\ref{H_gaussian}). The profile at the end-facet
(panel a) first transforms into a close-to-circular profile (panel c) followed by a hexagonal shape rotated by $\pi/6$ (panels d to f), a close-to-circular profile (panel g), and finally  a hexagonal shape
(panels h and i) with the same orientation as at the end-facet (panel a). 
Comparing with Fig.~\ref{fig2} this is qualitatively in excellent agreement with the experimental observations. The fact that the fully coherent scattering description qualitatively reproduces the experimentally observed
$\pi/6$ rotation gives strong indications of its nature; it is a phenomena caused by an interference between
the different gaussian elements used in the decomposition of the fundamental mode in the PCF. In Fig.~\ref{fig6}
we show the corresponding intensity distribution in the far-field limit which is in a very good agreement with the
experiments, see Fig.~\ref{fig3}. It is seen that the satellites are reproduced and are in fact oriented in the same way
as in the experiment. Moreover the relative intensities between the satellites and the main peak in Fig.~\ref{fig6} (left)
are very similar to the ones in Fig.~\ref{fig3} (left).

Finally, for fibers where the
air holes modify the overall gaussian profile sufficiently (not shown) we find indication of additional higher-order
spots further away from the center of the intensity distribution which can also be seen experimentally.

\section{Conclusion}

The evolution of the mode shape of a PCF with a triangular cladding has been investigated
in the transition from the near to the far field. When moving away from the near field at the focus of the fiber end-facet,
it has been observed that the hexagonal orientation is rotated two times by $\pi/6$ after which six satellites emerge in the
radiation pattern. In the far-field limit the satellites remain in the pattern, having a relative peak intensity more than
two orders of magnitude less than the main peak and with an orientation rotated by $\pi/6$ relative to the six inner holes
around the fiber core. All these observations have been reproduced theoretically, by approximating the near-field
distribution by a main gaussian peak from which six narrow gaussians located near the center of the six inner holes have
been subtracted. From the simulations it is concluded that the changes of shape in the radiation pattern are caused by an
interference between the different gaussian elements used in the decomposition of the fundamental mode in the PCF.

The results presented here are very important for understanding and analysing the behavior of the mode in many optical
systems based on photonic crystal fibers -- especially those involving imaging and focusing the mode. Furthermore, the
successful idea of decomposing the near field of the mode into seven localized distributions can be adapted in future work
aimed at a simple quantitative description of the near and far-field distributions, particularly for relating the measured
far field to the physical structure around the fiber core that influences the near field. The latter is very interesting
in the field of fiber measurement procedures, since the far-field analysis of conventional optical fibers cannot directly
be adapted to PCFs because of the lack of cylindrical symmetry.

\section*{Acknowledgments}
We thank J. Riishede and T.~P. Hansen (COM, Technical University of Denmark), P.~M.~W. Skovgaard and J. Broeng (Crystal Fibre A/S) for technical assistance and useful discussions.

\end{document}